\documentclass[iop]{emulateapj}

\usepackage{amsmath}
\usepackage{amssymb}
\usepackage{mathbbol}
\usepackage{dsfont}

\shorttitle{The VPOS Attains New Members}
\shortauthors{Pawlowski \& Kroupa}

\begin{document}

\title{The Vast Polar Structure of the Milky Way Attains New Members}

\author{Marcel S. Pawlowski}
\affil{Department of Astronomy, Case Western Reserve University,\\
              10900 Euclid Avenue, Cleveland, OH, 44106, USA}
\email{marcel.pawlowski@case.edu}
\and
\author{Pavel Kroupa}
\affil{Helmholtz-Institut f\"ur Strahlen- und Kernphysik, Rheinische Friedrich-Wilhelms-Universit\"at Bonn,\\
Nussallee 14-16, D-53115 Bonn, Germany}

\journalinfo{To be published in The Astrophysical Journal}
\submitted{Received 2014 April 22; accepted 2014 June 4}

\begin{abstract}
The satellite galaxies of the Milky Way (MW) align with and preferentially orbit in a vast polar structure (VPOS), which also contains globular clusters, stellar and gaseous streams. Similar alignments have been discovered around several other host galaxies. We test whether recently discovered objects in the MW halo, the satellite galaxy/globular cluster transition object named PSO J174.0675-10.8774 or Crater and three stellar streams, are part of the VPOS. Crater is situated close to the VPOS. Incorporating the new object in the VPOS-plane fit slightly improves the alignment of the plane with other features such as the Magellanic stream and the average orbital plane of the satellites co-orbiting in the VPOS. We predict Crater's proper motion by assuming that it, too, orbits in the VPOS. One of the three streams aligns well with the VPOS. Surprisingly, it appears to lie in the exact same orbital plane as the Palomar 5 stream and shares its distance, suggesting a direct connection between the two. The stream also crosses close to the Fornax dwarf galaxy and is oriented approximately along the galaxy's direction of motion. The two other streams cannot align closely with the VPOS because they were discovered in the direction of M31/M33, which is outside of the satellite structure. The VPOS thus attains two new members. This further emphasizes that the highly anisotropic and correlated distribution of satellite objects requires an explanation beyond the suggestion that the MW satellite system is an extreme statistical outlier of a $\Lambda$CDM sub-halo system.
\end{abstract}


\section{Introduction}

\begin{figure*}
   \centering
\includegraphics[scale=0.6]{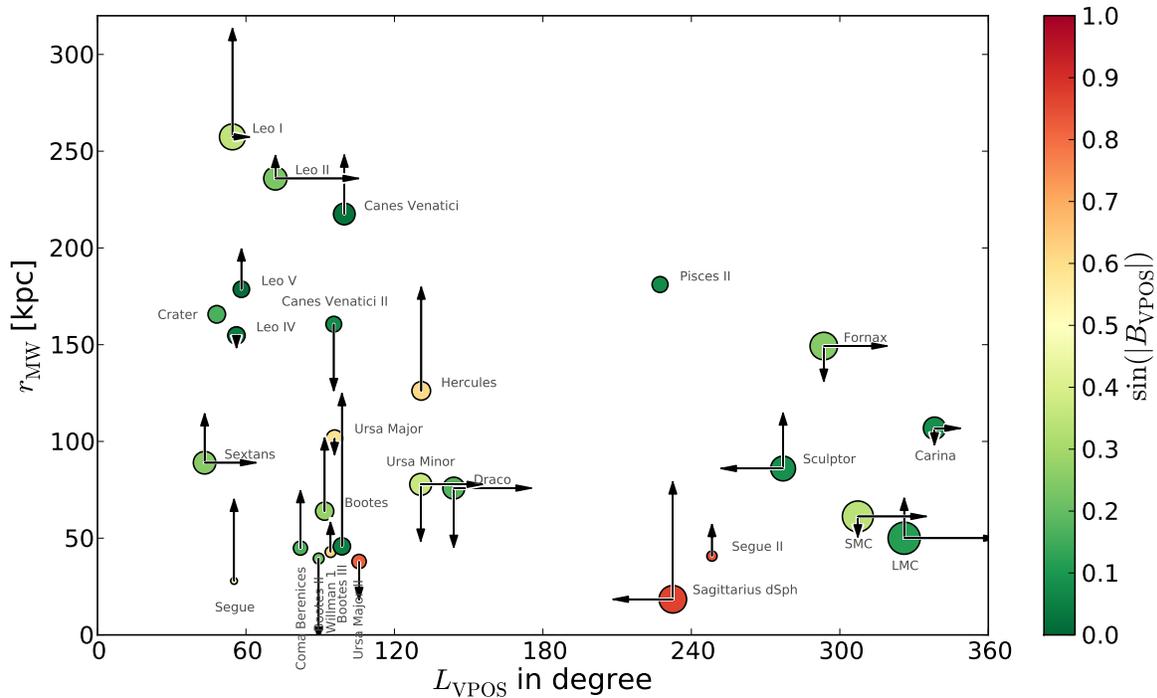}
   \caption{MW satellite galaxies in a VPOS-coordinate system. The angular position $L_{\mathrm{VPOS}}$\ along the VPOS-3 plane (measured from the intersection of the VPOS-3 plane with the MW plane) is plotted against Galactocentric distance $r_{\mathrm{MW}}$. The shading of the points indicates the angle $\sin(B_{\mathrm{VPOS}})$\ outside of the best-fit plane. More luminous objects are plotted with bigger symbols. The arrows indicate the amount of radial Galactocentric velocity (vertical, with 1.0\,km\,s$^{-1}$\ corresponding to 0.3\,kpc) and the direction and amount of the tangential velocity within the VPOS plane if a PM has been measured (horizontal, with 1.0\,km\,s$^{-1}$\ corresponding to $0.1^\circ$). Most MW satellites, in particular the more distant ones, are close to the VPOS-3, and most satellites with measured PMs orbit in the same direction within the plane (towards the right in the plot). The plotted data is compiled in Table \ref{tab:VPOScoord}. 
   }
              \label{fig:VPOScircleplot}
\end{figure*}

The distribution of satellite galaxies around the Milky Way (MW) is highly anisotropic \citep{LyndenBell1976,Kroupa2005,Kroupa2010,Pawlowski2013a}. They are arranged in a vast polar structure (VPOS), a narrow plane (root-mean-square height of 20 -- 30\,kpc, radius of 250\,kpc) almost perpendicular to the MW disk. The proper motions (PMs) of the 11 brightest MW satellite galaxies reveal that almost all of these orbit within the VPOS in the same direction, such that the satellite plane appears to be rotationally stabilised \citep{Pawlowski2013b}. In addition, globular clusters (GCs) classified as young halo (YH) objects \citep[e.g.][]{Mackey2005} define the same polar plane \citep{Majewski1994,Keller2012,Pawlowski2012a}.

Fig. \ref{fig:VPOScircleplot} shows the alignment of MW satellite galaxies by plotting their positions and velocities in a coordinate system motivated by the VPOS. The coordinate system has its origin in the Galactic center and its pole in the direction of the normal vector to the VPOS plane. We use the best-fit plane to the MW satellites excluding three outliers \citep[VPOS-3;][]{Pawlowski2013a} which has a normal vector pointing to $(l, b) = (169.5^\circ, -2.8^\circ)$. The VPOS longitude $L_{\mathrm{VPOS}}$\ is the angle measured along the best-fit plane and the VPOS latitude $B_{\mathrm{VPOS}}$\ measures the angle of a satellite away from the plane. The direction $(L_{\mathrm{VPOS}}, B_{\mathrm{VPOS}}) = (0.0^\circ, 0.0^\circ)$\ points toward $(l, b) = (259.5^\circ, 0.0^\circ)$\ in Galactic coordinates, which is the intersection of the VPOS-3 plane with the Galactic plane. From there $L_{\mathrm{VPOS}}$\ increases towards the Galactic North. This choice makes it aligned with the preferred orbital direction of satellites in the VPOS. Also plotted are the radial velocities of the satellites with respect to the Galactic center and, if PMs are available, the tangential velocity component projected into the VPOS plane. Almost all of these tangential velocities point to the right in the figure, a clear sign of the common orbital direction of the MW satellites in the VPOS. The data plotted in Fig. \ref{fig:VPOScircleplot} are collated in Table \ref{tab:VPOScoord}.

As satellite objects orbit around the MW, they may be disrupted by tidal forces acting on them and the material (stars or gas) they lose is spread out along streams. Many such streams have been discovered in the halo of the MW \citep[e.g.][]{Odenkirchen2003,Grillmair2006,Belokurov2006,Grillmair2009}, most of them thanks to the Sloan Digital Sky Survey (SDSS). These streams are also of importance with respect to the anisotropy of the MW satellites, because they reveal the orientation of the orbital plane of their progenitor. \citet{Pawlowski2012a} discovered that half of the then-known stellar and gaseous streams in the MW halo align with the VPOS, including the Magellanic Stream (MS). Thus these very different objects support the orbital alignment found using the satellite galaxy PMs.

A similar planar satellite galaxy structure, albeit consisting of only about half the satellite population, exists around the Andromeda galaxy M31 \citep{Ibata2013, Conn2013}. This Great Plane of Andromeda (GPoA) is seen edge-on from the MW and most of the satellite galaxies belonging to the GPoA share a common line-of-sight velocity trend, indicative of a co-orbiting satellite plane. \citet{Hammer2013} have pointed out that the GPoA also aligns with stellar structures in the halo of M31: the Giant Stream (GS), which also shares the common velocity trend seen in the satellite galaxies, and the North-West Stream 1 (NW-S1).

In addition to these satellite structures around the two major Local Group galaxies, alignments of several satellite galaxies, often also aligned with stellar and gaseous streams, have been found around more distant hosts \citep[e.g.][]{Galianni2010,Duc2011,Paudel2013,Karachentsev2014}. The fact that more than one satellite aligns with a stream in these cases indicates that these do not simply constitute objects embedded in their own tidal debris. Table \ref{tab:satellitestructures} compiles these currently known satellite structures. In addition, planar, linear and flattened alignments of more isolated dwarf galaxies have been discovered in the Local Group \citep{Pawlowski2013a,Pawlowski2014}, in its vicinity \citep{Bellazzini2013,Pawlowski2014} and in the M81 group \citep{Chiboucas2013}, respectively. Indeed, \citet{Chiboucas2013} write ''In review, in the few instances around nearby major galaxies where we have information, in every case there is evidence that gas poor companions lie in flattened distributions''

The existence of coherent satellite galaxy structures appears to be in conflict with current galaxy formation theories based on dark matter simulations \citep[e.g.][]{Kroupa2005,Metz2009,Pawlowski2012b,Pawlowski2013b,Ibata2014,Kroupa2014,Pawlowski2014b}. Alternative origins of the structures have been proposed, in particular that the satellites might be tidal dwarf galaxies (TDGs) instead of primordial dwarf galaxies embedded in dark matter sub-haloes \citep[e.g.][]{Kroupa1997,Metz2007b,Pawlowski2011,Casas2012,Pawlowski2012a,Hammer2013,Zhao2013}. Because this implies a different formation channel of the satellite galaxies, it is important to determine for each object in the halo of the MW whether it could be a members of the VPOS, since the properties of satellite objects within the structure might ultimately help to constrain its origin.

\begin{deluxetable*}{lccccccc}
\tabletypesize{\scriptsize}
\tablecaption{MW Satellites in VPOS-3 Coordinates \label{tab:VPOScoord}}
\tablehead{
\colhead{Name} & \colhead{$r_{\mathrm{MW}}$~(kpc)\tablenotemark{a}} & \colhead{$L_{\mathrm{VPOS}}~(^\circ)$\tablenotemark{b}} & \colhead{$\sin(B_{\mathrm{VPOS}})$\tablenotemark{c}} & \colhead{$v^{\mathrm{rad}}~(\mathrm{km\,s}^{-1})$\tablenotemark{d}} & \colhead{$v_{\mathrm{VPOS}}^{\mathrm{tan}}~(\mathrm{km\,s}^{-1})$\tablenotemark{e}}
& \colhead{$v_{\mathrm{VPOS}}^{\mathrm{perp}}~(\mathrm{km\,s}^{-1})$\tablenotemark{f}}
& \colhead{$L_V~(10^6 \mathrm{L}_{\sun})$}}
\startdata
Sagittarius dSph &   18 &  232 & 0.87 &  182 & -180 & -235 & 21.5\\ 
Segue &   28 &   55 & 0.59 &  120 & ---  & ---  & 0.0003\\ 
Ursa Major II &   38 &  106 & 0.82 &  -45 & ---  & ---  & 0.0041\\ 
Bootes II &   39 &   89 & 0.22 & -118 & ---  & ---  & 0.0010\\ 
Segue II &   41 &  248 & 0.85 &   34 & ---  & ---  & 0.0009\\ 
Willman 1 &   43 &   94 & 0.63 &   31 & ---  & ---  & 0.0010\\ 
Coma Berenices &   45 &   82 & 0.17 &   78 & ---  & ---  & 0.0037\\ 
Bootes III &   46 &   99 & 0.05 &  243 & ---  & ---  & 0.0171\\ 
LMC &   50 &  326 & 0.11 &   48 &  322 &  -12 & 1510\\ 
SMC &   61 &  307 & 0.34 &  -15 &  215 &  -39 & 461\\ 
Bootes &   64 &   92 & 0.28 &  106 & ---  & ---  & 0.0286\\ 
Draco &   76 &  144 & 0.18 &  -81 &  252 &   87 & 0.283\\ 
Ursa Minor &   78 &  131 & 0.37 &  -77 &  188 &   18 & 0.283\\ 
Sculptor &   86 &  277 & 0.09 &   75 & -189 &   71 & 2.29\\ 
Sextans &   89 &   43 & 0.26 &   64 &  145 &  180 & 0.437\\ 
Ursa Major &  102 &   96 & 0.59 &   -7 & ---  & ---  & 0.0139\\ 
Carina &  107 &  338 & 0.08 &   -8 &   43 &   72 & 0.377\\ 
Hercules &  126 &  131 & 0.61 &  158 & ---  & ---  & 0.0373\\ 
Fornax &  149 &  294 & 0.25 &  -40 &  194 &    3 & 20.3\\ 
Leo IV &  155 &   56 & 0.04 &   -0 & ---  & ---  & 0.0185\\ 
Canes Venatici II &  161 &   95 & 0.07 &  -94 & ---  & ---  & 0.0079\\ 
Crater &  166 &   48 & 0.16 & ---  & ---  & ---  & 0.0196\\ 
Leo V &  179 &   58 & 0.02 &   49 & ---  & ---  & 0.0108\\ 
Pisces II &  181 &  227 & 0.08 & ---  & ---  & ---  & 0.0086\\ 
Canes Venatici &  218 &  100 & 0.03 &   81 & ---  & ---  & 0.233\\ 
Leo II &  236 &   72 & 0.23 &   19 &  274 &  -62 & 0.738\\ 
Leo I &  257 &   55 & 0.35 &  166 &    7 &  153 & 5.50
\enddata
\tablecomments{MW satellite phase-space coordinates in VPOS-3 coordinates, as plotted in Fig. \ref{fig:VPOScircleplot}. These were calculated using the positions and line-of-sight velocities of the MW satellites as tabulated by \citet{McConnachie2012} and the proper motions as tabulated in \citet{Pawlowski2013b}.}
\tablenotetext{a}{Galactocentric distance.}
\tablenotetext{b}{Angle along the VPOS-3 plane as seen from Galactic center.}
\tablenotetext{c}{Angle out of the VPOS-3 plane as seen from Galactic center.}
\tablenotetext{d}{Galactocentric radial velocity.}
\tablenotetext{e}{Tangential velocity within VPOS-3 plane, sign indicates sense of rotation with positive being co-orbiting.}
\tablenotetext{e}{Velocity perpendicular to VPOS-3 plane.}
\end{deluxetable*}

\begin{deluxetable*}{lccccc}
\tabletypesize{\scriptsize}
\tablecaption{Known Correlated Dwarf Galaxy Structures \label{tab:satellitestructures}}
\tablehead{
\colhead{Host} & \colhead{Name} & \colhead{$N_{\mathrm{dwarf}}$\tablenotemark{a}} & \colhead{Kinematic Coherence\tablenotemark{b}} & \colhead{Aligned Streams\tablenotemark{c}} & \colhead{Reference}}
\startdata
            Milky Way & VPOS & $\geq 24$ &  Yes\tablenotemark{d} & Yes (stellar and gaseous, incl. MS)  &  1, 2 \\
            Andromeda & GPoA & $\geq 15$ &  Yes\tablenotemark{e} & Yes (stellar NW-S1 and GS) &  3, 4, 5\\
            NGC 1097 & Dog Leg & 2 & Unknown & Yes, stellar & 6\\
            NGC 5557 & Tidal Tail-E & 3 & Yes\tablenotemark{f} & Yes, stellar &  7, 8\\
            NGC 4216 & F1 & 3 &  Unknown & Yes, stellar &  9, 10\\
            NGC 4631 & Bridge & 3 &  Unknown & Possible stellar, H$\alpha$\ and HI bridge &  11\\
            \tableline
            M 81 group &   & 19 &  Unknown & Unknown\tablenotemark{g} &  12\\
            Local Group & NGC 3109 association & 5 & Yes\tablenotemark{h} & nNo stream known &  13, 14
\enddata
\tablenotetext{a}{Number of known dwarf galaxies aligning with the structure.}
\tablenotetext{b}{Do the dwarf galaxy velocities (if known) show a common trend?}
\tablenotetext{c}{Do known stellar or gaseous streams align with the structure?}
\tablenotetext{d}{8 of 11 brightest co-orbit in VPOS, one counter-orbits in the structure \citep{Pawlowski2013b}.}
\tablenotetext{e}{13 of 15 satellites follow a common velocity trend: the northern ones recede; southern ones approach in a M31 rest-frame \citep{Ibata2013}. The GS shares this velocity trend \citep{Hammer2013}.}
\tablenotetext{f}{Coherent velocity gradient along the three objects \citep{Duc2011}.}
\tablenotetext{g}{Flattening of gas-deficient dwarf galaxies along Supergalactic Plane, but more precise distances needed for three-dimensional analysis of the distribution \citep{Chiboucas2013}.}     
\tablenotetext{h}{Similar line-of-sight velocities placing them close to the MW at about the same time \citep{Pawlowski2014}.}
\tablerefs{
(1)~\citet{Pawlowski2012a}; (2) \citet{Pawlowski2013b}; (3) \citet{Ibata2013}; (4) \citet{Conn2013}; (5) \citet{Hammer2013}; (6) \citet{Galianni2010}; (7) \citet{Duc2011}; (8) \citet{Duc2014}; (9) \citet{Paudel2013}; (10) \citet{Martinez-Delgado2010}; (11) \citet{Karachentsev2014}; (12) \citet{Chiboucas2013}; (13) \citet{Bellazzini2013}; (14) \citet{Pawlowski2014}.
}
\end{deluxetable*}

We therefore test whether several recently discovered objects in the MW halo align with the VPOS. No such information is provided in the discovery papers. In Sect. \ref{sect:Crater} we will determine the alignment of the distant MW satellite object named PSO J174.0675-10.8774 or Crater \citep{Laevens2014,Belokurov2014}. In Sect. \ref{sect:Streams} we determine the orientation of three recently discovered stellar streams in the MW halo relative to the VPOS, other streams and the orbital poles of the MW satellites. These streams are the ATLAS stream \citep{Koposov2014}, the Pisces/Triangulum stream \citep{Bonaca2012,Martin2013} and the PAndAS MW stream \citep{Martin2014}. We end with concluding remarks in Sect. \ref{sect:Conclusion}.

To put the alignments into context, when discussing the positions and orientations in the following we also provide the probability $P_{\mathrm{random}}$. It either gives the probability that a randomly positioned object aligns to an angle of $\theta$\ or less (measured from the center of the MW) with the orientation of a given plane ($P_{\mathrm{random}} = P_{\mathrm{vector}} = \sin(\theta)$, the position vector has to be close to the plane), or the probability that two randomly oriented independent planes (such as the VPOS and a stream plane) are aligned to $\theta$\ or less ($P_{\mathrm{random}} = P_{\mathrm{planes}} = 1 - \cos(\theta)$, the normal vectors have to align). While it is not unlikely that a randomly positioned object is within $\theta = 30^{\circ}$\ of a given plane ($P_{\mathrm{vector}} = 0.5$), it is much less likely that a randomly oriented orbital plane aligns to $\theta \leq 30^{\circ}$\ ($P_{\mathrm{planes}} = 0.13$\footnote{Or half of this probability if the orbital direction is also required to agree with the preferred orbital direction of the satellites in the VPOS.}). 
An alignment of an object's orbital planes with the VPOS is thus much more informative than alignments in position alone, but the latter are of course a prerequisite for the former. For example, an object more distant than $40^{\circ}$\ from the VPOS cannot have an orbit which aligns to better than $40^{\circ}$\ with the VPOS. 
In the following, we will consider alignments to better than $\approx 30^{\circ}$\ to be consistent with VPOS-membership, because this is the typical scatter of observed orbital poles around their preferred direction \citep{Pawlowski2013b}, while two-thirds of the MW satellites positions even align to better than $20^{\circ}$\ with the VPOS-3 (see e.g. Table \ref{tab:VPOScoord}).


\section{Satellite object PSO J174.0675-10.8774/Crater}
\label{sect:Crater}

This object has been discovered independently in two different surveys. \citet{Laevens2014} found it in the photometric PAN-STARRS 3$\pi$\ survey and termed it PSO J174.0675-10.8774. They measure a heliocentric distance of $145 \pm 17$\,kpc using the red horizontal branch and determine the object's absolute magnitude to be $M_V = -4.3 \pm 0.2$, corresponding to a luminosity of $L_V = 1.1 \times 10^4 M_{\sun}$.
They also determine the object's structural parameters and argue that it is too round and compact to be consistent with the known MW dwarf galaxies, and that its metallicity and age are similar to those of known YH GCs in the outer MW halo. They interpret the object to be the most distant known MW GC.

Independently, \citet{Belokurov2014} found this object in the photometric VST ATLAS survey and termed it Crater. Using Red Clump stars they determine a larger heliocentric distance of 170\,kpc and a total absolute magnitude of $M_V = -5.5 \pm 0.5$ ($L_V = 1.4 \times 10^4 M_{\sun}$). They measure a larger half-light radius than \citet{Laevens2014}, 30\,pc compared to 20\,pc, and interpret several blue stars as blue loop stars, indicative of recent star formation, which is why they prefer to interpret the object as a dwarf galaxy. For reasons of readability, in the following we will adopt the name Crater for this object, irrespective of whether it is a GC or a dwarf galaxy.

While the object's structural properties do not allow a decisive categorization as an extended (young) GC or a compact dwarf galaxy, the exact classification is not essential for the following analysis. It can be expected that the object is part of the VPOS in either case, because that structure consists of both dwarf galaxies and YH GCs. Crater's estimated age of 8 to 10\,Gyr from \citet{Laevens2014} is consistent with the suggested formation time for the VPOS from the debris of a major galaxy encounter \citep{Pawlowski2012a,Zhao2013,Pawlowski2014}. Such events not only produced TDGs, but also super star clusters that might evolve into GCs \citep{Bournaud2008}. This is consistent with the similar planar alignment found for the YH GCs. Thus, in the TDG scenario the existence of transition objects between these two classes might not be too surprising.

For the following analysis we adopt Crater's position and distance modulus from table 1 in \citet{Belokurov2014}. We thus assume the larger of the two distance estimates, because this makes the plane offsets measured for Crater upper limits. Additionally, at the larger distance Crater has more impact on a plane-fit (only 5 of 27 known satellite galaxies are farther away).

Crater's position vector from the Galactic center is $18.9^{\circ}$\ inclined relative to the VPOSall ($P_{\mathrm{random}} = 0.32$\ if randomly positioned), the best-fitting plane to all 27 MW satellites considered in \citet{Pawlowski2013a}, while the absolute offset from that plane is 65\,kpc. It is only $9.5^{\circ}$\ ($P_{\mathrm{random}} = 0.16$) away from the VPOS-3, the plane fit excluding three apparently unrelated satellite outliers \citep{Pawlowski2013a}, which aligns better with the average orbital plane of the MW satellites co-orbiting in the VPOS \citep{Pawlowski2013b}. Crater is offset by $39.1 \pm 1.3$\,kpc from this VPOS-3 plane. If Crater is added as a satellite to that plane fit, the offset reduces to $31.4 \pm 0.8$\,kpc, approximately 1.5 times the best-fit plane's root-mean-square height\footnote{If Crater is at a distance of 145\,kpc, these offsets are reduced to $57.4 \pm 2.1$\,kpc from the VPOSall, $34.7 \pm 1.3$\,kpc from VPOS-3, and $29.1 \pm 0.8$\,kpc from VPOS-3 if Crater is included in the plane fit.}. The plane properties (compiled in Table \ref{tab:VPOSfits}) are very similar to those excluding Crater (compare to table 3 of \citealt{Pawlowski2013a}), the most distinct change is that the new plane fits align 3 to $4^{\circ}$\ more closely with the average orbital plane than those excluding Crater. The new VPOS-3 plane normal is only $4^{\circ}$\ away from the orbital pole of the Large Magellanic Cloud ($P_{\mathrm{random}} = 0.002$) and the new VPOSall normal coincides to better than $2^{\circ}$\ ($P_{\mathrm{random}} = 0.0006$) with the average direction of the stream normals (Sect. \ref{sect:Streams}).

\begin{deluxetable}{lcc}
\tabletypesize{\scriptsize}
\tablecaption{VPOS Parameters with Crater Included in the Satellite Sample.\label{tab:VPOSfits}}
\tablehead{
\colhead{Name} & \colhead{VPOSall (+ Crater)} & \colhead{VPOS-3 (+Crater)}
}
\startdata
$n \begin{pmatrix} l \\ b \end{pmatrix}$\ ($^{\circ}$)\tablenotemark{a} & $\begin{pmatrix}  159.3 \\ -2.9 \end{pmatrix}$  & $\begin{pmatrix}  172.6 \\ -2.3 \end{pmatrix}$ \\ 
$D_{\mathrm{MW}}$\ (kpc)\tablenotemark{b} & $ 7.2 \pm 0.3$ & $ 10.1 \pm 0.2$ \\ 
$\Delta$\ (kpc)\tablenotemark{c} & $30.9 \pm 0.4$ & $20.7 \pm 0.3$ \\ 
$c/a$\tablenotemark{d} & $0.318 \pm 0.004$ & $0.217 \pm 0.002$
\enddata
\tablenotetext{a}{Direction of best-fitting plane's normal vector in Galactic coordinates.}
\tablenotetext{b}{Offset of best-fitting plane from MW center.}
\tablenotetext{c}{Root-mean-square height of satellites above best-fit plane.}
\tablenotetext{d}{Short-to-long ($c/a$) and intermediate-to-long ($b/a$) root-mean-square axis ratios.}
\end{deluxetable}

Eight of the 11 most luminous MW satellites are consistent with co-orbiting in a common plane within the uncertainties, as can be seen from their very concentrated distribution of orbital poles \citep[directions of angular momenta around the MW center,][]{Pawlowski2013b}. Crater is only $4.5^{\circ}$\ ($P_{\mathrm{random}} = 0.08$) away from the plane defined by the average orbital pole of the six best-aligned co-orbiting MW satellites (and $13.8^{\circ}$\ from that of the eight best-aligned; $P_{\mathrm{random}} = 0.24$) and therefore it is consistent with being a member of this co-orbiting structure of MW satellites. Knowledge of Crater's PM would allow a more decisive constraint on its (orbital) alignment with the VPOS.

\begin{figure}
   \centering
\includegraphics[scale=0.6]{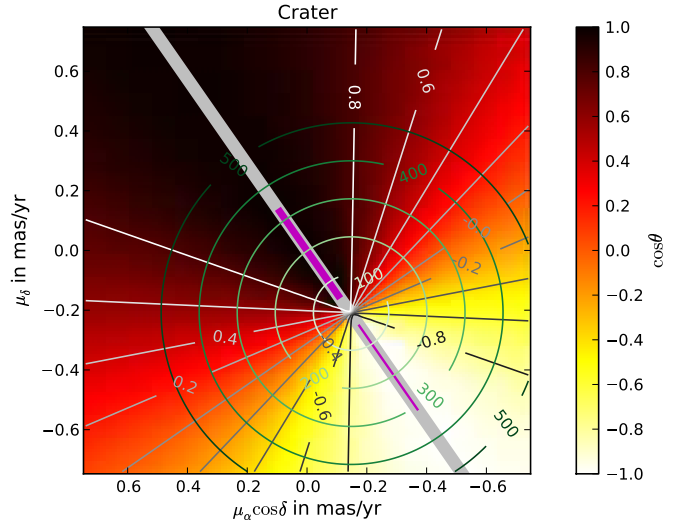}
   \caption{Predicted PM of Crater assuming it moves within the VPOS-3. For each combination of the two PM components ($\mu_{\alpha} \cos \delta$\ and $\mu_{\delta}$) the map illustrates the angle $\theta$\ between the VPOS-3 plane (without Crater) and the orbital plane which would result from the PM. The radial gray contour lines also illustrate this angle, measured in $\cos \theta$. 
   PM with $\cos \theta > 0.8$\ ($\cos \theta < -0.8$) result in orbital planes aligning to better than $37^{\circ}$\ with the VPOS-3 and are therefore co-orbiting (counter-orbiting). 
   The circular contours (green in the online journal) indicate the absolute speed of Crater relative to the MW in $\mathrm{km\,s}^{-1}$, assuming that its heliocentric line-of-sight velocity is 150\,km\,s$^{-1}$. 
   The broad light gray line marks PMs resulting in the best alignment of the satellite galaxy's orbital plane and the VPOS-3. 
   The thick (thin) line on top of the broad light gray one (magenta in the online journal) indicates the predicted PM if the satellite is co-orbiting (counter-orbiting). 
\label{fig:propmo}}
    \end{figure}

No PM has yet been measured for Crater, but the object is closely aligned with the VPOS-3 and the average orbital plane. We can therefore use the method presented in \citet{Pawlowski2013b} to predict its PM, assuming that it either co- or counter-orbits within the satellite plane. Because no line-of-sight velocity to Crater has been measured yet we assume that its heliocentric radial velocity is 150\,km\,s$^{-1}$, from averaging the radial velocities of Leo IV and Leo V that are both very close to Crater (as discussed by \citealt{Belokurov2014}). This choice of radial velocity does not affect the orientation of the predicted PM line but only constrains the maximum and minimum predicted PM to values such that Crater remains bound to the MW. This results in assumed minimum and maximum absolute Galactocentric three-dimensional velocities of 50 and 317\,km\,s$^{-1}$. Under these assumptions the PM of Crater will be close to the line defined by
$$
\begin{pmatrix} \mu_{\alpha} \cos \delta \\ \mu_{\delta} \end{pmatrix}_{\mathrm{co}} = 
\begin{pmatrix} -0.10 \\ -0.15 \end{pmatrix}
~\mathrm{to}~
\begin{pmatrix} +0.09 \\ +0.13 \end{pmatrix}\,\mathrm{mas\,yr}^{-1}
$$
if it is co-orbiting close to the VPOS-3 or 
$$
\begin{pmatrix} \mu_{\alpha} \cos \delta \\ \mu_{\delta} \end{pmatrix}_{\mathrm{counter}} = 
\begin{pmatrix} -0.17 \\ -0.25 \end{pmatrix}
~\mathrm{to}~
\begin{pmatrix} -0.37 \\ -0.53 \end{pmatrix}\,\mathrm{mas\,yr}^{-1}
$$
if it is counter-orbiting like Sculptor. The predictions for Crater's PM are also shown in Fig. \ref{fig:propmo}.

\citet{Belokurov2014} mention the direction of the pole of a great circle connecting the three satellite galaxies Leo IV, Leo V and Crater, which are within $10^{\circ}$\ of each other and at similar Galactocentric distances of 155, 180 and 170 (or 145) kpc, respectively. In Galactic coordinates the pole of this great circle points to $(l, b) = (208.6^{\circ}, -20.0^{\circ})$\ (magenta cross in Fig. \ref{fig:ASP}). This is $\approx 30^{\circ}$\ ($P_{\mathrm{random}} = 0.13$) away from the average orbital pole of the eight best-aligned orbital poles, thus still compatible with a common orbital plane of all three satellites aligned roughly with the VPOS.


\section{Streams}
\label{sect:Streams}

   \begin{figure*}
   \centering
\includegraphics[scale=0.4]{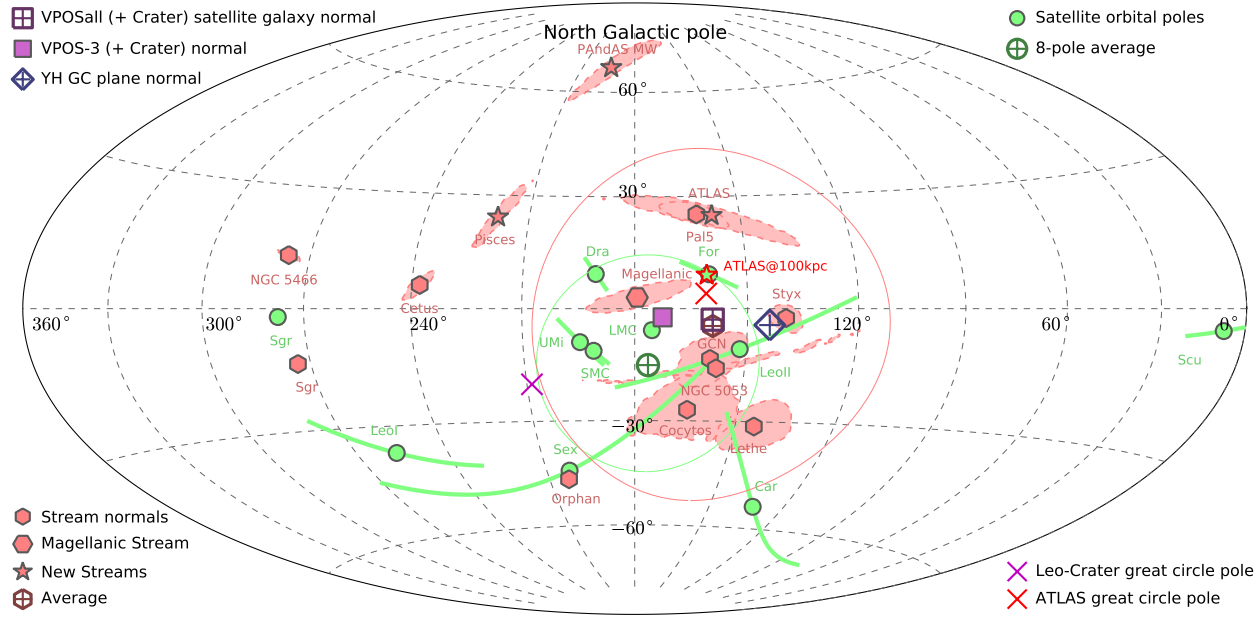}
   \caption{All-sky plot showing the orientation of the planes fitted to the positions of the satellite galaxies (squares, magenta in the online journal) and young halo globular clusters (YH GC, diamond, blue in the online journal), the orientation of individual satellite orbital planes (circles, green in the online journal; with uncertainty lines) and of individual streams (hexagons, red in the online journal; with $1\sigma$\ uncertainty contours). Note that it is not positions that are shown, but the directions of plane-normal vectors, orbital poles and stream normals, i.e. vectors perpendicular to these features. Axial directions (all normal vectors, but not the orbital poles which indicate the vectorial angular momentum directions) have only been plotted in the range $120^{\circ} < l < 300^{\circ}$, i.e. the mirrored directions were omitted for clarity. 
   See the text for a more detailed description and discussion.
   }
              \label{fig:ASP}
    \end{figure*}

Following the method presented in \citet{Pawlowski2012a}, we determine the stream normal vectors for three new MW streams. A stream normal describes the orientation of the orbital plane of a stream, which is the plane containing two anchor points on the stream and the Galactic center. We assume that the Galactic center is at a distance of 8.3\,kpc from the Sun \citep{McMillan2011}.

Fig. \ref{fig:ASP} shows the resulting stream-normal directions together with the VPOS plane-normal directions and satellite orbital poles. For this plot only long streams which cover at least $5^{\circ}$\ on the sky are considered, because shorter streams do not reliably trace the orbital plane of a stream. As was shown by \citet{Pawlowski2012a}, streams close to the Sun are biased to lying outside of the VPOS. We therefore only consider streams with heliocentric distances of at least 10\,kpc. The normal directions of the three streams discovered since the original analysis of \citet{Pawlowski2012a} are indicated with a star symbol in Fig. \ref{fig:ASP}, while the others are indicated by hexagons. The filled contours denote the $1\sigma$\ uncertainty region of the stream normals, determined with a Monte Carlo method by varying the stream anchor point positions $10^4$~times within their uncertainties.

The directions of the normal vectors to the VPOSall and VPOS-3 from Table \ref{tab:VPOSfits} are indicated with a large square with a plus sign and with a filled light square in Fig. \ref{fig:ASP}, respectively. In addition, the plot shows the orbital poles (directions of angular momenta) for the 11 brightest MW satellites and their uncertainties as the dots and great-circle segments, respectively \citep{Pawlowski2013b}. Also plotted, as a blue diamond with a plus sign, is the direction of the normal to the plane fitted to the YH GCs from \citet{Pawlowski2013a}.

The average direction of all stream normals shown in Fig. \ref{fig:ASP} is the open hexagon with a plus sign. It points to $(l, b) = (159.2^{\circ}, -4.6^{\circ})$, around which they have a spherical standard distance of $47.9^{\circ}$\ (thin circle). This direction is almost identical to the normal of the plane fitted to all known MW satellite galaxies and Crater (VPOSall). Even though one direction is determined from satellite positions while the other is defined by the orientations of streams from disrupted objects, both describe a very similar orientation. Eight of the 14 streams in this plot align to $\lesssim 30^{\circ}$\ with the VPOSall ($P_{\mathrm{random}} = 0.13$\ for an individual stream, probability of 0.015\% to have 8 out of 14 randomly oriented streams align this well with an independently defined direction).

The excellent alignment of Sextans' most-likely orbital pole and the Orphan stream normal is probably coincidental, because the part of the stream close to Sextan's position is considerably closer to the Sun \citep[$\approx 20$\,kpc;][]{Newberg2010} than the satellite galaxy \citep[$\approx 90$\,kpc;][]{Lee2009} and the galaxy's orbital pole is very uncertain due to its uncertain PM.

In the following we discuss the adopted data for the new streams and their orientations. For more information on the other streams we refer the reader to the discussion in Sect. 3.2 of \citet{Pawlowski2012a}.

\subsection{ATLAS Stream}
\label{sect:ATLAS}

This stream has been discovered by \citet{Koposov2014} in the first data release of the VST ATLAS survey. Its narrow width suggests a GC origin.

\citet{Koposov2014} report that a great circle with a pole at $(\alpha,\delta) = (77^{\circ},47^{\circ})$\ aligns well with the stream on the sky. This corresponds to $(l,b) = (161^{\circ},4^{\circ})$\ in Galactic coordinates (red cross in Fig. \ref{fig:ASP}), which is close to the VPOS-3 normal direction and very close to the VPOSall. As seen from the Sun, this stream thus aligns closely with the VPOS.

According to \citet{Koposov2014}, the stream's signal-to-noise in a background-subtracted density map is optimized if a distance of 20\,kpc is adopted. Consequently parallax effects cannot be ignored because the stream does not orbit around the Sun but around the Galactic center. To determine the stream normal's orientation we follow the method presented in \citet{Pawlowski2012a}. We estimate the stream anchor point positions from fig. 1 of \citet{Koposov2014} to be $(\alpha, \beta)_{\mathrm{start}} = (18^{\circ},-25^{\circ})$\ and $(\alpha, \beta)_{\mathrm{end}} = (30^{\circ},-32^{\circ})$. The uncertainty in the anchor-point positions is assumed to be $1^{\circ}$. The distance to both anchor points is first assumed to be 20\,kpc, while we adopted a distance uncertainty of $\pm 2.5$\,kpc for each anchor point individually.

The parallax effect moves the stream normal a bit away from the VPOS pole (toward the north in Fig. \ref{fig:ASP}). The resulting stream normal direction, the vector perpendicular to the plane described by the two anchor points and the Galactic center, points to $(l,b) = (158.1^{\circ},24.8^{\circ})$. The ATLAS stream is almost polar with respect to the MW disk and is aligned with the VPOS to $\approx 30^{\circ}$\ ($P_{\mathrm{random}} = 0.13$). Increasing the adopted stream anchor point distance improves the alignment with the VPOS again, the stream normal moves toward the pole of the great circle defined by the stream on the sky (red cross in Fig. \ref{fig:ASP}), with which it would coincide if the distance were infinite.

Interestingly the new ATLAS stream has almost the same orientation as the well-known Palomar 5 stream, as revealed by the match of its stream normal with that of Palomar 5 pointing to $(l,b)_{\mathrm{Pal5}} = (162^{\circ},25^{\circ})$\ and the overlap of both stream normal's uncertainty contours. If the distance estimate to the ATLAS stream by \citet{Koposov2014} is correct, it is possible that both streams orbit in exactly the same plane, and in addition both streams would have a very similar Galactocentric distance of $\approx 20$\,kpc. Furthermore, Palomar 5's metallicity is [Fe/H] = -1.41 \citep{Harris2010}. Independently of this, \citet{Koposov2014} argue that such a metallicity is consistent with the noisy color-magnitude diagram of the ATLAS stream when discussing a possible association of the stream with the $\approx 90^{\circ}$-distant GC Pyxis. However, Palomar 5 is about $140^{\circ}$\ away from the center of the observed ATLAS stream segment. Streams of similar angular extent are the more nearby $\> 80^{\circ}$\ long and narrow GD-1 stream \citep{Grillmair2006b,Carlberg2013} and the wider Sagittarius stream, which wraps at least once around the MW \citep[e.g.][]{Fellhauer2006,Penarrubia2010,Law2010,PilaDiez2014}, but it will require numerical models to test if and under which conditions Palomar 5 might have formed such a long stream. 
If the ATLAS stream were a distant part of the Palomar 5 stream, this would increase that streams's known extent to more than $140^{\circ}$\ on the sky and thus provide extremely detailed constraints for the modelling of Palomar 5's orbit, which would be very helpful for attempts to determine the MW potential from fitting stellar streams \citep[e.g.][]{Koposov2010, Sanders2013, Bovy2014}.

Another possible association can be suspected from the stream's closeness to the dSph MW satellite Fornax (see fig. 1 in \citealt{Koposov2014}). Curiously, this apparent alignment is not only present in the position on the sky. As can be seen in Fig. \ref{fig:ASP}, the ATLAS stream normal and the orbital pole of Fornax, situated at $(l,b)_{\mathrm{pole}} = (160^{\circ}, 9^{\circ})$ \citep{Pawlowski2013b}, are only $\approx 15^{\circ}$\ ($P_{\mathrm{random}} = 0.034$) apart. Thus Fornax orbits the MW approximately in the plane defined by the stream. This suggest a possible scenario in which the ATLAS stream's progenitor was a GC that once belonged to Fornax. However, Fornax is at a distance of $\approx 150$\,kpc, much farther away than the stream according to the estimate by \citet{Koposov2014}. If instead the stream were at a distance of $\approx 100$\,kpc, its stream normal (open red star in Fig. \ref{fig:ASP}) and the orbital pole of Fornax would match up perfectly. 

Fornax is co-orbiting in the VPOS, while averaging the three available PM measurements for Palomar 5 \citep{Palma2002,Scholz1998,Schweitzer1993} indicates that this GC is counter-orbiting, more like the dSph satellite Sculptor. Therefore one of these two possible associations could be ruled out if the direction of motion of the ATLAS stream can be determined.

\subsection{Pisces/Triangulum Stream}
\label{sect:Pisces}

This stream has been discovered by both \citet{Bonaca2012}, who named it the Triangulum stream, and \citet{Martin2013}, who named it the Pisces stream. The same stream has been detected in the PAndAS survey \citep{Martin2014}. Due to its narrow width it probably originates from a GC. We use the stream's start and end-positions reported by \citet{Martin2013} as our stream anchor points: $(\alpha, \beta)_{\mathrm{start}} = (21^{\circ},35^{\circ})$\ and $(\alpha, \beta)_{\mathrm{end}} = (24^{\circ},23^{\circ})$. The uncertainty in the anchor-point positions is assumed to be $1^{\circ}$. The distance to both anchor points is assumed to be 35\,kpc, with an adopted distance uncertainty of $\pm 3$\,kpc, again adopted from \citet{Martin2013}. This places the most likely stream normal at $(l,b) = (219.0^{\circ}, 24.2^{\circ})$, $64^{\circ}$\ ($P_{\mathrm{random}} = 0.56$) away from the VPOSall and $52^{\circ}$\ ($P_{\mathrm{random}} = 0.38$) away from the VPOS-3. It does not align with the VPOS.

\subsection{PAndAS MW Stream}
\label{sect:pandas}

This stream has been discovered in the deep photometric PAndAS survey centered on M31 by \citet{Martin2014}, who estimate its width to be several hundred parsec, such that its most likely progenitor is a dwarf galaxy. We estimate the stream anchor points from fig. 2 of \citet{Martin2014} to be $(\xi, \nu)_{\mathrm{start}} = (-9^{\circ},6^{\circ})$\ and $(\xi, \nu)_{\mathrm{end}} = (10^{\circ},-1^{\circ})$\ in the M31-centric coordinates $(\xi, \nu)$. To convert these to equatorial coordinates $(\alpha, \beta)$, we add the position of M31 $(\alpha, \delta)_{\mathrm{M31}} = (10.7^{\circ}, 41.3^{\circ})$, to these: $(\alpha, \beta) = (\xi, \nu) + (\alpha, \delta)_{\mathrm{M31}}$. We adopt $1^{\circ}$\ uncertainties in the anchor-point positions. Following \citet{Martin2014} we adopt an average distance to the stream of 17\,kpc. However, to incorporate the likely distance gradient along the stream reported by \citet{Martin2014}, we assume the start point to be 1.5\,kpc more distant than this and the end point to be 1.5\,kpc closer. The distance uncertainty is assumed to be $\pm 3$\,kpc for both anchor points independently.

The most likely stream pole then is $(l,b)_{\mathrm{pole}} = (193.6^{\circ}, 67.7^{\circ})$, $75^{\circ}$\ ($P_{\mathrm{random}} = 0.74$) away from the VPOSall and $72^{\circ}$\ ($P_{\mathrm{random}} = 0.69$) away from the VPOS-3. The PAndAS MW stream is therefore almost co-planar with the disk of the MW, as already noted by \citet{Martin2014}. The stream does not align with the VPOS, which is not surprising because the direction to M31 points $\approx 40^{\circ}$\ ($P_{\mathrm{random}} = 0.64$) away from the VPOS \citep{Pawlowski2013a}, such that any stream discovered in this direction must be outside of the structure.


\section{Discussion and Conclusion}
\label{sect:Conclusion}

We have tested whether recently found stellar systems in the MW halo align with the VPOS consisting of satellite galaxies, GCs and streams \citep{Pawlowski2012a}.

The MW satellite object PSO J174.0675-10.8774 / Crater \citep{Laevens2014,Belokurov2014}, sharing properties with both dSph satellite galaxies and GCs, is situated close to the VPOS and can therefore be considered to be another member of this structure. We provide updated fits to the VPOS-planes in Table \ref{tab:VPOSfits}. We have furthermore predicted the object's PM if it orbits within the VPOS, as has been empirically found for most of the MW satellites for which PMs have been measured \citep{Pawlowski2013b}.

The Pisces/Triangulum stream \citep{Bonaca2012,Martin2013} and the PAndAS MW stream \citep{Martin2014} do not align with the VPOS, as was to be expected because they both have been found in the direction of M31, outside of the VPOS. 

The ATLAS stream \citep{Koposov2014} aligns well with the VPOS. In addition to the previously suggested possible association of the stream with the GC Pyxis, our work reveals two other possible connections. The stream might be related to the GC Palomar 5 and its stream (due to the almost perfect alignment of both stream's planes, similar distances, and consistent metallicity) or to a GC that once belonged to the MW satellite galaxy Fornax (due to the very similar orientation of the satellite galaxy's orbital plane and the stream). The latter scenario would probably require that the stream is much more distant than currently estimated and will possibly be ruled out if deeper observations of the stream confirm its closeness. A determination of the stream's orbital direction would also be able to decide between these two scenarios.

Whether one of these alignments is due to a real association remains to be seen. The fact that most satellite objects in the MW align with and move within the VPOS certainly increases the chances of finding apparent alignments. However, the common alignment along the VPOS itself already suggests a causal connection or common origin in a more fundamental sense. Similar correlated structures of dwarf galaxies are known around an increasing number of host galaxies (as compiled in Table \ref{tab:satellitestructures}). This emphasizes that the satellite system of the MW (and that of M31) cannot be an extreme statistical outlier from on average more isotropic satellite systems as expected from cosmological simulations \citep{Pawlowski2014b}. Instead, observations appear to suggest that correlated satellite systems might even be the norm, with the corresponding implications for cosmology \citep{Kroupa2012a,Kroupa2012b,Kroupa2014}.




\end{document}